\documentclass[a4paper,aps,prd,10pt,preprintnumbers,showpacs,twocolumn,superscriptaddress,nofootinbib,amsmath,amssymb]{revtex4-1}\def\PRDstyle#1{#1}\def\JCAPstyle#1{}\def\Abstract#1{\begin{abstract}#1\end{abstract}}
\usepackage[utf8]{inputenc}
\usepackage[T1]{fontenc}
\usepackage{cmap}
\usepackage{ulem}

\def\Order#1{{\cal O}\left(#1\right)}

\begin{document}
\title{Blandford-Znajek mechanism in the general stationary axially-symmetric black-hole spacetime}

\JCAPstyle{
\author[\star]{R. A. Konoplya,}\emailAdd{roman.konoplya@gmail.com}
\author[\diamond]{J. Kunz,}\emailAdd{jutta.kunz@uni-oldenburg.de}
\author[\diamond\ddagger]{A. Zhidenko}\emailAdd{alexander.zhidenko@uni-oldenburg.de}
\affiliation[\star]{Research Centre for Theoretical Physics and Astrophysics,\\ Institute of Physics, Silesian University in Opava,\\ Bezručovo náměstí 13, CZ-74601 Opava, Czech Republic}
\affiliation[\diamond]{Institute of Physics, University of Oldenburg, D-26111 Oldenburg, Germany}
\affiliation[\ddagger]{Centro de Matemática, Computação e Cognição (CMCC),\\ Universidade Federal do ABC (UFABC),\\ Rua Abolição, CEP: 09210-180, Santo André, SP, Brazil}
\arxivnumber{2102.10649}
}

\PRDstyle{
\author{R. A. Konoplya}\email{roman.konoplya@gmail.com}
\affiliation{Research Centre for Theoretical Physics and Astrophysics, Institute of Physics, Silesian University in Opava, Bezručovo náměstí 13, CZ-74601 Opava, Czech Republic}
\author{J. Kunz}\email{jutta.kunz@uni-oldenburg.de}
\affiliation{Institute of Physics, University of Oldenburg, D-26111 Oldenburg, Germany}
\author{A. Zhidenko}\email{alexander.zhidenko@uni-oldenburg.de}
\affiliation{Institute of Physics, University of Oldenburg, D-26111 Oldenburg, Germany}
\affiliation{Centro de Matemática, Computação e Cognição (CMCC), Universidade Federal do ABC (UFABC),\\ Rua Abolição, CEP: 09210-180, Santo André, SP, Brazil}
}

\Abstract{
We consider the Blandford-Znajek process of electromagnetic extraction of energy from a general axially symmetric asymptotically flat slowly rotating black hole. Using the general parametrization of the black-hole spacetime we construct formulas for the flux of the magnetic field and the rate of energy extraction, which are valid not only for the Kerr spacetime, but also for its arbitrary axially symmetric deformations. We show that in the dominant order these quantities depend only on a single deformation parameter, which relates the spin frequency of a black hole with its rotation parameter.
}

\maketitle

\section{Introduction}
{Magnetic fields in the vicinity of black holes have been actively studied in a great number of works  (see for example \cite{Ammon:2011je,Turimov:2019afv,Wu:2015fwa,Yang:2014zva,Junior:2021dyw,Santos:2021nbf,Wang:2016wcj,Dvornikov:2021hps,Allahyari:2019jqz,Frolov:2014zia,Konoplya:2008hj,Konoplya:2006gg,Konoplya:2007yy,Konoplya:2020xam} and references therein), because they drive various processes of radiation and particles motion.
Among a number of mechanisms of extracting rotational energy from black holes, such as the Penrose process \cite{Penrose}, superradiance \cite{Starobinsky:1973aij,Starobinskil:1974nkd}, and others, it is the Blandford-Znajek process \cite{Blandford:1977ds}, which is characterized by high efficiency and could be realized in an astrophysical environment.
If a rotating black hole is immersed in an external magnetic field, a Lorentz transformation says that an electric field appears in the co-rotating frame which induces a separation of charges, that is, an electric current in the inertial frame. This way the rotational energy of a black hole is transferred into the energy of the currents outside the black hole. This is the essence of the Blandford-Znajek process \cite{Kinoshita:2017mio}, which could explain the gamma-ray bursts \cite{Lee:1999se,Lloyd-Ronning:2019ncg}.
In a stricter sense, if the force-free magnetosphere is established as a result of the equation of state for the magnetized equatorial accretion disks, then the interaction between the magnetosphere and the black hole's ergosphere leads to extraction of the rotational energy \cite{Grignani:2018ntq,Tursunov:2019oiq}.

This process has been extensively studied during the past four decades in the Kerr background, that is, for an axially symmetric and asymptotically flat black hole in the Einstein theory of gravity. At the same time, there are a number of reasons to consider various, generally speaking not small \cite{Konoplya:2016pmh}, deviations from Kerr geometry, either because of the modification of the Einstein theory of gravity or due to taking into consideration some tidal forces in the vicinity of the black hole. Therefore, it would be interesting to see whether the Blandford-Znajek process can be described in terms of the most general deformations of Kerr spacetime  and whether there are some general characteristics of this process, which are independent on the background geometry. An attempt in this direction was made in \cite{Pei:2016kka}. However, there were considered only particular deformations given by the so called Johannsen metric \cite{Johannsen:2015pca}. This metric introduces some ad hoc deformations, however, it does not represent the general form of the axially symmetric asymptotically flat black hole in a parametrized form and, therefore, one cannot judge about general properties of black holes within this approach. On the contrary, the general parametrization for axially symmetric asymptotically flat black holes, which is suitable for any metric theory of gravity, was suggested in \cite{Konoplya:2016jvv}. This parametrization has superior convergence and strict hierarchy of parameters and, therefore, frequently allows one to describe essential astrophysically relevant properties of black holes in terms of only a few parameters of deformation \cite{Konoplya:2020hyk,Konoplya:2021slg}.

Here we consider the Blandford-Znajek process in the background of this arbitrary parametrized axially symmetric and asymptotically flat black hole \cite{Konoplya:2020hyk}. Although part of our calculations is performed for arbitrary rotation, the full solution of the equations of electrodynamics can be found in analytic form only perturbatively, in the regime of slow rotation. In this regime we show that the rate of energy extraction for the split monopole magnetic field remarkably depends only on its flux and the black hole's spin frequency, i.~e. can be described by a single parameter of deviation from the Kerr spacetime. Notice that in this approach, only the rotation, but not the deformation parameter, must be small. In other words, we propose a universal description of the Blandford-Znajek process for arbitrary axially symmetric black holes in any metric theory of gravity in the regime of slow rotation.

The paper is organized as follows. In Sec.~\ref{sec:gen} we briefly outline the general parametrization of an axially symmetric black hole spacetime developed in \cite{Konoplya:2016jvv}.
In Sec.~\ref{sec:equatorial} we write out the values of the coefficients of the parametrization in the equatorial plane.
Sec.~\ref{sec:BZ} is devoted to description of the Blandford-Znajek process in the low rotation regime for a general axisymmetric black holes and also with a particular example of Kerr solution.
Finally, in the Conclusions, we summarize our results and their implications for alternative theories of gravity.

\section{General rotating black hole}\label{sec:gen}

A general stationary axisymmetric (rotating) black hole in an arbitrary metric theory of gravity can be represented by the following line element \cite{Konoplya:2016jvv}
\begin{eqnarray}\label{genmetric}
ds^2&=&-\dfrac{N^2(r,\theta)-W^2(r,\theta)\sin^2\theta}{K^2(r,\theta)}dt^2\\\nonumber
&& - 2W(r,\theta)r\sin^2\theta dtd\varphi+K^2(r,\theta)r^2\sin^2\theta d\varphi^2
\PRDstyle{\\\nonumber&&}
+\Sigma(r,\theta)\left(\dfrac{B^2(r,\theta)}{N^2(r,\theta)}dr^2 + r^2d\theta^2\right)\,,
\end{eqnarray}
with
\begin{equation}
\label{Sigmacondition}
\Sigma(r, \theta) \equiv 1+\frac{a^2\cos^2\theta}{r^2}\,,
\end{equation}
where $a$ is the rotation parameter defined through the asymptotic mass $M$ and angular momentum $J$ as follows,
$$a\equiv\frac{J}{M}.$$
Within this ansatz the coordinate system is fully fixed in such a way that the coordinates $r$ and $\theta$ are mutually orthogonal and orthogonal to the coordinates $t$ and $\phi$, associated with the Killing vectors. Thus, the black-hole metric is fully determined by the four metric functions $N$, $B$, $W$ and $K$ of $r$ and $\theta$.

It is important to emphasise that we assume the spacetime to be stationary and axisymmetric everywhere outside the horizon. Thus, there are two Killing vectors, spacelike and timelike, associated with the coordinates $\varphi$ and $t$ respectively. This assumption excludes some rather exotic models of rotating black holes, such as disformed Kerr black holes \cite{Anson:2020trg}. The disformed black holes, supported by a time-depending scalar field, provide an example of a non-circular geometry, which can appear in a higher order scalar-tensor theory of gravity \cite{BenAchour:2020fgy}. The Killing vector, which is timelike in the far region, becomes a spacelike one in some vicinity of the black hole, which is causally connected to the asymptotic region. Since the one-form associated to the Killing vector is not integrable, the resulting geometry is non-circular \cite{Anson:2020trg}. At the same time, in that case there is a region where no stationary observer can exist, which obviously breaks our assumptions. Thus, we leave the consideration of the rotating black holes with non-circular geometry beyond the scope of our work.

The event horizon is defined by the equation
\begin{equation}\label{horizon}
 N^2(r,\theta)=0,
\end{equation}
and the ergoregion corresponds to
\begin{equation}
 0<N^2(r,\theta)<W^2(r,\theta).
\end{equation}

It is convenient to define
\begin{eqnarray}
x\equiv1-\frac{r_0}{r},
\qquad
y\equiv\cos\theta,
\end{eqnarray}
where $r_0$ is the event horizon radius in the equatorial plane ($\theta=\pi/2$, $y=0$).

The parametrization consists in a double expansion to obtain a generic axisymmetric metric expression: a continued-fraction expansion in terms of a compact radial coordinate and a Taylor expansion in terms of the cosine of the polar angle for the polar dependence. These choices lead to a superior convergence in the radial direction and to the exact limit in the equatorial plane.
Thus, the metric functions are represented as series expansion in terms of the polar coordinate $y$,
\begin{equation}\label{yexp}
\begin{array}{rcl}
N^2 &=& xA_0(x)+\sum\limits_{i = 1}^{\infty}A_i(x)y^i\,, \\
& &\\
B &=& 1+\sum\limits_{i = 0}^{\infty}B_i(x)y^i\,,\\
& &\\
W &=& \sum\limits_{i = 0}^{\infty}\dfrac{W_i(x)y^i}{\Sigma}\,, \\
& &\\
K^2 &=& 1+\dfrac{a W}{r}+\sum\limits_{i = 0}^{\infty}\dfrac{K_i(x)y^i}{\Sigma}\,,
\end{array}
\end{equation}
where we introduced the following functions:
\begin{subequations}
\label{fdef}
\begin{eqnarray}
\label{bdef}
B_i(x) &=& b_{i0}(1-x)+{\tilde B}_i(x)(1-x)^2\,,
\\ \nonumber \\
\label{wdef}
W_i(x) &=& w_{i0}(1-x)^2+{\tilde W}_i(x)(1-x)^3\,,
\\ \nonumber \\
\label{kdef}
K_i(x) &=& k_{i0}(1-x)^2+{\tilde K}_i(x)(1-x)^3\,,
\\ \nonumber \\
\label{a0def}
A_0(x) &=& 1-\epsilon_0(1-x)+(a_{00}-\epsilon_0+k_{00})(1-x)^2
\PRDstyle{\nonumber \\ && \phantom{1}}
+ {\tilde A}_0(x)(1-x)^3\,,
\\ \nonumber \\
\label{aidef}
A_{i>0}(x) &=& K_i(x)+\epsilon_{i}(1-x)^2+a_{i0}(1-x)^3+
\PRDstyle{\nonumber \\ && \phantom{K_i(x)}}
+ {\tilde A}_i(x)(1-x)^4\,.
\end{eqnarray}
\end{subequations}

The functions ${\tilde A}_i(x)$, ${\tilde B}_i(x)$, ${\tilde W}_i(x)$, and ${\tilde K}_i(x)$ are defined via continued fractions in terms of the compact radial coordinate $x$,
\begin{subequations}
\label{tiltedfunctions}
\begin{eqnarray}
{\tilde A}_i(x)& = &\dfrac{a_{i1}}{1+\dfrac{a_{i2}x}{1+\dfrac{a_{i3}x}{1+\ldots}}}\,,
\end{eqnarray}\begin{eqnarray}
{\tilde B}_i(x)& = &\dfrac{b_{i1}}{1+\dfrac{b_{i2}x}{1+\dfrac{b_{i3}x}{1+\ldots}}}\,,
\end{eqnarray}\begin{eqnarray}
{\tilde W}_i(x)& = &\dfrac{w_{i1}}{1+\dfrac{w_{i2}x}{1+\dfrac{w_{i3}x}{1+\ldots}}}\,,
\end{eqnarray}\begin{eqnarray}
{\tilde K}_i(x)& = &\dfrac{k_{i1}}{1+\dfrac{k_{i2}x}{1+\dfrac{k_{i3}x}{1+\ldots}}}\,,
\end{eqnarray}
\end{subequations}
where the coefficients are determined by comparing the series expansions near the event horizon ($x=0$).
On the other hand, the coefficients $\epsilon_i$, $a_{i0}$, $b_{i0}$, $w_{i0}$, $k_{i0}$ for $i = 0,1,2,3\ldots$ are fixed in order to match the asymptotic behavior near spatial
infinity ($x=1$).

This way the general parametrization of the black-hole spacetime is similar in the spirit to the parametrized post-Newtonian (PPN) formalism, satisfies the PPN asymptotics in the far region, but is valid in the whole space outside the black hole. The above general parametrization and its spherically symmetric limit have been recently applied to construction of analytical approximations for various numerical black-hole solutions \cite{Konoplya:2020jgt,Hennigar:2016gkm,Kokkotas:2017zwt,Kokkotas:2017ymc,Konoplya:2018arm,Konoplya:2019fpy,Konoplya:2019goy}.

\section{Equatorial-plane coefficients}\label{sec:equatorial}

There are a few reasons to consider the values of the coefficients of the parametrization in the equatorial plane. First of all, relations between some
characteristics such as, the surface gravity, spin frequency, asymptotic mass and rotation parameter do not depend on the point in which they are found, so that the easiest way is to relate these characteristics on the equatorial plane and further use them for the whole space. The second observation is related to the slow rotation regime.
It turns out that in this case the dominant contributions are coming mainly from the values of the coefficients on the equatorial plane. This should not come as a surprise, because the expansion in the polar direction is made around the equatorial plane and the slow rotation is associated with it, while for the faster rotation more terms of expansion in powers of $y\equiv\cos\theta$ around the equatorial plane should be taken into account.

The asymptotic coefficients in the equatorial plane are
\begin{itemize}
\item $\epsilon_0$, which relates the asymptotic mass and the horizon radius,
\begin{equation}\label{epsilon0}
\epsilon_0 = \frac{2M-r_0}{r_0}\,,
\end{equation}
\item $a_{00}$ and $b_{00}$, which depend on the post-Newtonian parameters $\beta$ and $\gamma$,
\begin{eqnarray}
\label{a00}
a_{00}& = &(\beta-\gamma)\frac{2M^2}{r_0^2} =
\frac{(\beta-\gamma)(1+\epsilon_0)^2}{2}\,,
\\
\label{b00}
b_{00}& = &(\gamma-1)\frac{M}{r_0} = \frac{(\gamma-1)(1+\epsilon_0)}{2}\,,
\end{eqnarray}
\item and $k_{00}$ is fixed as follows,
\begin{equation}\label{k00}
k_{00} = \frac{a^2}{r_0^2}\,.
\end{equation}
\end{itemize}

In addition we shall consider the Lense-Thirring-like term, which is defined by the asymptotic parameter $w_{00}$,
\begin{equation}\label{w00}
w_{00}=\frac{2J}{r_0^2} = \frac{2Ma}{r_0^2} = \frac{J}{M^2}\frac{(1+\epsilon_0)^2}{2}\,.
\end{equation}

We can also express some near-horizon characteristics in terms of the equatorial-plane coefficients.
The black-hole spin frequency $\Omega_H$ is a constant at the event horizon
\begin{equation}\label{OmegaH}
  \Omega_H=\frac{g^{t\phi}}{g^{tt}} \Biggr|_{g^{rr}=0}=-\frac{g_{t\phi}}{g_{\phi\phi}} \Biggr|_{g^{rr}=0}=\frac{W(r,\theta)}{rK^2(r,\theta)} \Biggr|_{N^2(r,\theta)=0}\!\!\!\!\!\!\!\!\!\!\!\!\!\!\!\!.
\end{equation}
In particular, one can calculate $\Omega_H$ by taking its value in the equatorial plane,
\begin{eqnarray}\label{OmegaH-equatorial}
\Omega_H&=&\frac{W_0(0)}{r_0+aW_0(0)+r_0K_0(0)}
\PRDstyle{\\\nonumber&=&}
\JCAPstyle{=}
\frac{2Ma+w_{01}r_0^2}{r_0(r_0^2+a^2)+2Ma^2+aw_{01}}.
\end{eqnarray}
Therefore, one can interpret $w_{01}$ as a parameter of deviation of the black hole's spin frequency from its Kerr value (see Appendix~\ref{sec:Kerr}).

The event horizon is a Killing horizon, corresponding to the Killing vector
\begin{equation}\label{Killing}
  \xi^{\mu}=(1,0,0,\Omega_H), \qquad \xi^{\mu}\xi_{\mu}\Biggr|_{N^2(r,\theta)=0}=0.
\end{equation}

The surface gravity at the event horizon,
\begin{equation}
\kappa_g=\sqrt{\frac{1}{2}(\nabla_{\nu}\xi^{\mu})(\nabla_{\mu}\xi^{\nu})},
\end{equation}
is finite iff \cite{Zaslavskii:2018lbb}
\begin{equation}
\frac{\partial N^2}{\partial\theta}=0.
\end{equation}
Therefore, we conclude that, for the physically relevant black-hole spacetimes, the event horizon is located at the constant radial coordinate $r=r_0$, i.~e. $r=r_0$ is a solution of (\ref{horizon}) in our coordinates for any $\theta$,
$$N^2(r_0,\theta)=0.$$

After some algebra we obtain for the surface gravity
\begin{equation}\label{surface-gravity}
\kappa_g=\frac{1}{2B(r_0,\theta)\sqrt{\Sigma(r_0,\theta)K^2(r_0,\theta)}}\frac{\partial N^2(r_0,\theta)}{\partial r},
\end{equation}
which is also a constant (does not depend on $\theta$).

Again, considering $\theta=\pi/2$, we can express the surface gravity in terms of the equatorial-plane coefficients,
\begin{eqnarray}\label{surface-gravity-horizon}
\kappa_g&=&\frac{A_0(0)}{(2+2B_0(0))\sqrt{r_0^2+ar_0W_0(0)+r_0^2K_0(0)}}
\\\nonumber
&=&\frac{1+(a^2/r_0^2)-2\epsilon_0+a_{00}+a_{01}}{2(1+b_{00}+b_{01})\sqrt{r_0^2+ar_0w_{01}+2a^2+a^2\epsilon_0}},
\end{eqnarray}
where we have used that \cite{Konoplya:2016jvv}
\begin{equation}\label{k01}
  k_{01}=0.
\end{equation}

With the above relations at hand we are ready to consider the parametrized metric functions in the slow rotation regime.

First we notice that, due to the symmetry of the line element with respect to substitution
$$\phi\to-\phi, \qquad J\to-J,$$
the functions $A_i(x)$, $B_i(x)$, $K_i(x)$, and $W_i(x)/a$ depend on $a^2$. Moreover, since in the slow-rotation regime, the functions $N^2$, $B$, $K^2$, and $W$ do not depend on the polar variable, we conclude that
$$A_i(x)=\Order{a}^2, \quad B_i(x)=\Order{a}^2, \quad K_i(x)=\Order{a}^2,$$
and
$$W_{i>0}=\Order{a}^3, \quad W_0(x)=\Order{a}.$$

Therefore, by definition (\ref{OmegaH-equatorial}),
\begin{equation}
\Omega_H=\Order{a}.
\end{equation}

Hence it follows that
\begin{eqnarray}\nonumber
\Sigma\phantom{^2}&=&1+\Order{a}^2=1+\Order{\Omega_H}^2,\\\nonumber
N^2 &=& xA_0(x) + \Order{a}^2= xA_0(x) + \Order{\Omega_H}^2,\\\nonumber
B\phantom{^2} &=& 1+B_0(x) + \Order{a}^2 = 1+B_0(x) + \Order{\Omega_H}^2,\\\label{slow-rotation}
K^2 &=& 1+ \Order{a}^2 = 1+ \Order{\Omega_H}^2,\\\nonumber
W\phantom{^2} &=& W_0(x)+ \Order{a}^3 = W_0(x)+ \Order{\Omega_H}^3.
\end{eqnarray}
Note that the first relation follows from the definition (\ref{Sigmacondition}).

\section{Blandford–Znajek mechanism}\label{sec:BZ}

In the original paper of Blandford and Znajek \cite{Blandford:1977ds} the energy- and momentum- extraction from the Kerr black hole surrounded by a stationary, axisymmetric, force-free, magnetized plasma was studied using the Boyer-Lindquist coordinates. However, in these coordinates the electromagnetic field requires appropriate boundary condition at the event horizon because the metric coefficient $g_{rr}$ in the Boyer-Lindquist diverges there, leading to a singular point in the Maxwell equations. Here we shall follow the approach of McKinney and Gammie \cite{McKinney:2004ka}, who used the Kerr-Schild coordinates. Since the corresponding line element is regular at the horizon, there is no need to specify boundary conditions at this point. For the generic axisymmetric black-hole metric we introduce the ingoing Eddington–Finkelstein-like variables,
\begin{equation}\label{EFcoord}
\begin{array}{rcl}
 d\tau &=& dt+C(r,\theta)dr,\\
 d\phi &=& d\varphi+C(r,\theta)\dfrac{W(r,\theta)}{rK^2(r,\theta)}dr,
\end{array}
\end{equation}
where
$$C(r,\theta)=\sqrt{\Sigma(r,\theta)K^2(r,\theta)}\dfrac{B(r,\theta)}{N^2(r,\theta)}.$$

In terms of the 1-forms (\ref{EFcoord}) the line element (\ref{genmetric}) takes the following form
\begin{eqnarray}\label{EFmetric}
ds^2&=&-\dfrac{N^2(r,\theta)-W^2(r,\theta)\sin^2\theta}{K^2(r,\theta)}d\tau^2\\\nonumber
&& - 2W(r,\theta)r\sin^2\theta d\tau d\phi+K^2(r,\theta)r^2\sin^2\theta d\phi^2\\\nonumber
&&+\Sigma(r,\theta)r^2d\theta^2+2B(r,\theta)\sqrt{\dfrac{\Sigma(r,\theta)}{K^2(r,\theta)}}drd\tau\,.
\end{eqnarray}

Notice that $\tau(t,r,\theta)$ and $\phi(\varphi,r,\theta)$ are not smooth functions of the radial and polar coordinates.
Because we assume that the black hole has a regular horizon, it is possible to prove that there are always smooth coordinate transformations $\tau(t,r,\theta)$ and $\phi(\varphi,r,\theta)$, such that the resulting metric tensor is regular at the event horizon (see Appendix~\ref{sec:regularEF}).
However, since no function depends on $\tau$ and $\phi$, in this section we shall use the metric (\ref{EFmetric}) without loss of generality.

The electromagnetic field tensor in a force-free magnetosphere satisfies \cite{Okamoto}
\begin{equation}\label{force-free}
  F_{\mu\nu}J^{\nu}=0,
\end{equation}
where $J^{\nu}$ is the current and
$$F_{\mu\nu}=\dfrac{\partial A_{\mu}}{\partial x^{\nu}}-\dfrac{\partial A_{\nu}}{\partial x^{\mu}}$$
is the Faraday tensor.
Assuming that the electromagnetic field depends on $\theta$ and $r$, this tensor takes the following form \cite{McKinney:2004ka}
\begin{equation}\label{Faraday}
  F_{\mu\nu} = \sqrt{-g}
\left(\begin{array}{cccc}
0 & - \omega B^{\theta} & \omega B^r & 0 \\
\omega B^{\theta} & 0 & B^{\phi} & - B^{\theta} \\
- \omega B^r & -B^{\phi} & 0 & B^r \\
0 & B^{\theta} & -B^r & 0
\end{array} \right),
\end{equation}
where $g$ is the determinant of the metric tensor (\ref{EFmetric})
$$g=-B^2(r,\theta)\Sigma^2(r,\theta)r^4\sin^2\theta.$$

Now we are in a position to calculate energy and angular momentum transport. Let $T_{\mu\nu}$ be the total energy-momentum tensor,
\begin{equation}\label{energy-momentum tensor}
  T_{\mu}^{\nu}=F_{\mu\lambda}F^{\nu\lambda}-\frac{1}{4}\delta_{\mu}^{\nu}F_{\kappa\lambda}F^{\kappa\lambda}, \qquad \nabla_{\nu}T_{\mu}^{\nu}=0.
\end{equation}
Then the conserved electromagnetic energy and angular momentum flux are \cite{Blandford:1977ds}
\begin{equation}\label{fluxdef}
  {\cal E}^{\nu}\equiv-T_{\tau}^{\nu}, \qquad {\cal L}^{\nu}\equiv T_{\phi}^{\nu}.
\end{equation}

Substituting (\ref{Faraday}) and (\ref{EFmetric}) into (\ref{energy-momentum tensor}), for the radial fluxes we have
\begin{eqnarray}
{\cal E}^{r}&=&\left(F_{\tau\theta} F_{\theta\phi}g^{r\phi} - F_{r\theta} F_{\tau\theta} g^{rr} - F_{\tau\theta}^2 g^{\tau r}\right)g^{\theta\theta},\\
{\cal L}^{r}&=&\left(F_{\theta\phi}^2 g^{r\phi} - F_{r\theta}F_{\theta\phi}g^{rr} - F_{t\theta}F_{\theta\phi} g^{\tau r}\right)g^{\theta\theta}.
\end{eqnarray}
Finally we obtain
\begin{eqnarray}\nonumber
  {\cal E}^{r}&=&\omega {\cal L}^{r}=N^2(r,\theta)r^2 \sin^2\theta\cdot \omega(r,\theta)\Biggl(-B^r(r,\theta)B^{\phi}(r,\theta)
\\\label{flux}
&&+(B^r(r,\theta))^2\cdot C(r,\theta)\left(\frac{W(r,\theta)}{rK^2(r,\theta)}-\omega(r,\theta)\right)\Biggr).
\end{eqnarray}

The energy flux from the black hole is given by (\ref{flux}) taken on the event horizon
\begin{equation}
F_E(\theta)\equiv {\cal E}^{r}(r_0,\theta),
\end{equation}
and the total rate of energy extraction from the black hole is given by the integral over the event horizon surface
\begin{equation}
  P_{BZ}=\iint \sqrt{-g}F_E(\theta)d\theta d\phi = 4\pi\intop_0^{\pi/2}\sqrt{-g}F_E(\theta)d\theta.
\end{equation}
Similarly, for the momentum extraction rate from (\ref{flux}) we obtain,
\begin{equation}
  M_{BZ}=4\pi\intop_0^{\pi/2}\sqrt{-g}{\cal L}^{r}(r_0,\theta)d\theta=4\pi\intop_0^{\pi/2}\sqrt{-g}\frac{F_E(\theta)}{\omega(r_0,\theta)}d\theta.
\end{equation}

Taking into account (\ref{OmegaH}) and (\ref{horizon}), we obtain
\begin{equation}\label{FE}
  F_E(\theta)=\omega(\Omega_H-\omega)(B^r)^2\cdot B\sqrt{\Sigma K^2}\cdot r^2 \sin^2\theta \Biggr|_{r=r_0}.
\end{equation}
This a generalization of the Blandford-Znajek formula for the rate of energy extraction from an arbitrary axisymmetric black hole.

It is interesting to note that, although the expression for the energy flux (\ref{FE}) depends explicitly on the metric function $B(r_0,\theta)$, the total energy extraction in terms of the magnetic field strength and spin frequency depends only on $K^2(r_0,\theta)$ at the horizon
\begin{equation}\label{FEinv}
\sqrt{-g}F_E(\theta)=\omega(\Omega_H-\omega)(F_{\theta\phi})^2\cdot \sqrt{\frac{K^2(r_0,\theta)}{\Sigma(r_0,\theta)}}\cdot \sin\theta,
\end{equation}
while the other metric coefficients affect (\ref{FEinv}) through the magnetic field $F_{\theta\phi}$, which is a solution of the Maxwell equations in the vicinity of the black-hole.

This leads to the universal behaviour of the energy extraction rate in the slow-rotating regime. It was shown in \cite{Tchekhovskoy:2009ba} that, for the split monopole magnetic field,
\begin{equation}\label{Br}
  B^r(r_0,\theta)=B^r_0+\Order{\Omega_H}^2.
\end{equation}
Then, taking into account (\ref{slow-rotation}), the total magnetic field flux is given by the following expression\footnote{Notice that by definition $\Phi$ differs from \cite{Tchekhovskoy:2009ba} by the factor 2.}
\begin{eqnarray}\nonumber
\Phi&=&\iint \sqrt{-g}B^{r}d\theta d\phi=4\pi\intop_0^{\pi/2}B^{r}\cdot B\cdot\Sigma\cdot r^2\sin\theta d\theta\\
&=&4\pi B^r_0 r_0^2(1+B_0(0))+\Order{\Omega_H}^2.
\end{eqnarray}
The maximum rate of energy extraction is achieved for $\omega=\Omega_H/2$,
\begin{eqnarray}\nonumber
  P_{BZ}&=&4\pi\intop_0^{\pi/2}\frac{\Omega_H^2}{4}(B^r_0)^2(1+B_0(0))^2r_0^4\sin^3\theta d\theta+\Order{\Omega_H}^4\!\!\!\!\\
\label{PBZ}
&=&\frac{\Omega_H^2}{6\pi}\cdot\left(\frac{\Phi}{2}\right)^2+\Order{\Omega_H}^4.
\end{eqnarray}

Thus, we conclude that, in the slow-rotating regime, the energy extraction rate depends on the black-hole geometry only through the value of $\Omega_H$ (\ref{OmegaH-equatorial}). Thus, for slow rotation, the difference in the Blandford-Znajek effect for various black-hole spacetimes will be stipulated by the single deformation coefficient $w_{01}$, which is the deviation of the spin frequency from its Kerr value. This dependence has been discussed for two different deformations given by the Johannsen metric \cite{Pei:2016kka}, which does not represent the most general rotating black-hole geometry. We have shown that for an arbitrary axisymmetric black hole in a metric theory of gravity the maximum energy extraction rate in the slow rotating regime is the same (\ref{PBZ}).

For faster rotation the near-horizon geometry also manifests through the magnetic-field configuration, which is a solution of the Maxwell equation in the particular curved spacetime, which can be obtained numerically or perturbatively in terms of $\Omega_H$ \cite{Tchekhovskoy:2009ba}.

\section{Conclusions}

The magnetic field in the vicinity of a black hole is an important constituent of its environment which affects processes of radiation and particles motion \cite{Ammon:2011je,Turimov:2019afv,Wu:2015fwa,Yang:2014zva,Junior:2021dyw,Santos:2021nbf,Wang:2016wcj,Dvornikov:2021hps,Allahyari:2019jqz,Frolov:2014zia,Konoplya:2008hj,Konoplya:2006gg,Konoplya:2007yy,Konoplya:2020xam}.
The Blandford-Znajek process \cite{Blandford:1977ds} is an astrophysically relevant mechanism of transformation of the black-hole rotational energy into the electromagnetic one, which was suggested as a central engine for gamma-ray bursts \cite{Lee:1999se}. This process has been extensively studied for the Kerr spacetime, but not for alternative theories of gravity. Here we have studied the Blandford-Znajek process using the general parametrization of the axially-symmetric and asymptotically flat black holes in the regime of slow rotation as the background. It turns out that, for the slow rotation, the qualitative characteristics of the process, such as the rate of extraction of energy and momentum, depend only on the three parameters: mass, angular momentum, and a single deformation parameter -- the deviation of the black hole's spin frequency from its Kerr value $w_{01}$. This way we have found universal relations for various quantities describing the Blandford-Znajek process of axially-symmetric black holes in arbitrary metric theories of gravity in the regime of slow rotation. Our work could, in principle, be continued to the higher orders of expansion in the polar direction, allowing to describe the fast rotation regime. In a similar fashion, using the general parameterized description \cite{Bronnikov:2021liv} it would also be interesting to know whether there is any mechanism of extracting of rotational energy for wormholes.

\begin{acknowledgments}
R.~A.~K. acknowledges the support of the grant 19-03950S of Czech Science Foundation (GAČR).
A.~Z. was supported by the Alexander von Humboldt Foundation, Germany.
J.~K. acknowledges support by the DFG Research Training Group 1620 ``Models of Gravity''
and the COST Actions CA15117 and CA16104.
\end{acknowledgments}

\appendix
\section{Kerr black hole}\label{sec:Kerr}
For the Kerr black hole the only nonzero asymptotic coefficients are \cite{Konoplya:2016jvv}
\begin{eqnarray}
\epsilon_0&=&\frac{a^2}{r_0^2}\,,\\
w_{00}&=&\frac{a}{r_0}+\frac{a^3}{r_0^3}\,,\\
k_{00}&=&\frac{a^2}{r_0^2}\,,\\
a_{20} &=& \frac{a^2}{r_0^2}+\frac{a^4}{r_0^4}\,.
\end{eqnarray}

Therefore, from (\ref{epsilon0}) it follows that the mass is given by
$$M=\frac{r_0^2+a^2}{2r_0};$$
(\ref{a00}) and (\ref{b00}) imply that the post-Newtonian parameters $\beta=\gamma=1$.

The nonzero near-horizon coefficients are
\begin{eqnarray}
a_{21} &=& -\frac{a^4}{r_0^4}\,,\\
k_{21} &=& -\frac{a^2}{r_0^2}\,,\\
k_{22} &=& -\frac{a^2}{r_0^2}\,,\\
k_{23} &=& \frac{a^2}{r_0^2}\,.
\end{eqnarray}

Since $w_{01}=0$, (\ref{OmegaH-equatorial}) reads
\begin{equation}\label{OmegaH-Kerr}
  \Omega_H^{(Kerr)}=\frac{a}{r_0^2+a^2}=\frac{a}{2Mr_0}.
\end{equation}

Taking into account that $a_{01}=b_{01}=0$, from (\ref{surface-gravity-horizon}) we find the surface gravity,
\begin{equation}\label{surface-gravity-Kerr}
\kappa_g^{(Kerr)}=\frac{r_0^2-a^2}{2r_0(r_0^2+a^2)}=\frac{r_0-M}{2Mr_0}.
\end{equation}

Finally, substituting the coefficients into (\ref{yexp}), we obtain the explicit form for the metric functions,
\begin{equation}\label{yexpKerr}
\begin{array}{rcl}
N^2(r,\theta) &=& \dfrac{(r-r0)(rr_0-a^2)}{r_0r^2}\,, \\
& &\\
B(r,\theta) &=& 1\,,\\
& &\\
W(r,\theta) &=& \dfrac{a}{r_0}\cdot\dfrac{r_0^2+a^2}{r^2+a^2\cos^2\theta}\,, \\
& &\\
K^2(r,\theta) &=& 1+\dfrac{a^2}{r^2}+\dfrac{a^2(a^2+r_0^2)\sin^2\theta}{rr_0(r^2+a^2\cos^2\theta)}\,.
\end{array}
\end{equation}

Using the explicit expressions (\ref{yexpKerr}), we find that for the Kerr black hole (\ref{FEinv}) takes the form (cf. (4.5) of \cite{Blandford:1977ds})
\begin{equation}\label{FEKerr}
  F_E(\theta)=\omega(\Omega_H-\omega)\left(\frac{F_{\theta\phi}(r_0,\theta)}{r_0^2+a^2\cos^2\theta}\right)^2\cdot (r_0^2+a^2).
\end{equation}

\section{Regular Eddington–Finkelstein-like coordinates}\label{sec:regularEF}
In order to simplify calculations in Sec.~\ref{sec:BZ} we have used the variables, $\tau$ and $\phi$, defined through nonclosed 1-forms (\ref{EFcoord}). This choice of coordinates does not affect the results because all the metric functions in (\ref{EFmetric}) do not depend on $\tau$ and $\phi$. However, for the sake of completeness, we show here that one can define the Eddington–Finkelstein-like coordinates, in which the event horizon is a regular point. For this purpose we explicitly define the transformations
\begin{equation}\label{regular-coord}
\begin{array}{rcl}
\tau&=&t+T(r,\theta),\\
\phi &=& \varphi+\Psi(r,\theta),
\end{array}
\end{equation}
where $T(r,\theta)$ and $\Psi(r,\theta)$ are smooth functions for $r>r_0$ and
\begin{equation}\label{regular-def}
\begin{array}{rcl}
\dfrac{\partial T}{\partial r}&=&C(r,\theta), \\ \\
\dfrac{\partial \Psi}{\partial r}&=&C(r,\theta)\dfrac{W(r,\theta)}{rK^2(r,\theta)}.
\end{array}
\end{equation}
From Eqs.~(\ref{surface-gravity}) and (\ref{OmegaH}) it follows that the righthand sides of Eq.~(\ref{regular-def}) do not depend on $\theta$ in the vicinity of the event horizon $r=r_0$,
\begin{equation}\label{regular-transform}
\begin{array}{rcl}
\dfrac{\partial T}{\partial r}&=&\left(\dfrac{1}{2\kappa_g}+\Order{r-r_0}\right)\times\dfrac{\partial\ln N^2}{\partial r}, \\ \\
\dfrac{\partial \Psi}{\partial r}&=&\left(\dfrac{\Omega_H}{2\kappa_g}+\Order{r-r_0}\right)\times\dfrac{\partial\ln N^2}{\partial r}.
\end{array}
\end{equation}

In the limit of the Kerr black hole Eqs.~(\ref{regular-transform}) coincide with the transformation from the Boyer-Lindquist to the Kerr-Schild map at the horizon (cf.~(5) and~(6) of~\cite{McKinney:2004ka}). However, the coordinates $\tau$ and $\phi$ are not reduced to the Kerr-Schild ones in the Kerr limit, because Eqs.~(\ref{regular-transform}) do not coincide with the Kerr-Schild transformation matrix for $r>r_0$. For an arbitrary rotating black hole, the coordinates satisfying (\ref{regular-transform}) penetrate the future horizon, being regular in its vicinity. This can be shown by constructing the following expansion
\begin{equation}\label{regular-exp}
\begin{array}{rcl}
T(r,\theta)&=&\dfrac{1}{2\kappa_g}\ln N^2(r,\theta)+T_0(\theta)\PRDstyle{\\&&\!\!\!\!\!\!}+T_1(\theta)(r-r_0)+T_2(\theta)(r-r_0)^2+\ldots,\\
\Psi(r,\theta)&=&\dfrac{\Omega_H}{2\kappa_g}\ln N^2(r,\theta)+\Psi_0(\theta)\PRDstyle{\\&&\!\!\!\!\!\!}+\Psi_1(\theta)(r-r_0)+\Psi_2(\theta)(r-r_0)^2+\ldots.
\end{array}
\end{equation}

The functions $T_1(\theta),T_2(\theta),\ldots$, and $\Psi_1(\theta),\Psi_2(\theta),\ldots$ depend on the near-horizon parameters of the black hole defined via (\ref{tiltedfunctions}). Their closed form can be obtained by substituting (\ref{regular-exp}) into (\ref{regular-transform}). In particular, we have
\begin{eqnarray}\nonumber
  T_1(\theta)&=&\frac{1}{4\kappa_g}\frac{\partial}{\partial r}\ln\left(\frac{\Sigma(r,\theta) K^2(r,\theta) B^2(r,\theta)}{(\partial N^2(r,\theta)/\partial r)^2}\right)\Biggr|_{r=r_0},
  \\\nonumber
  \Psi_1(\theta)&=&\frac{\Omega_H}{4\kappa_g}\frac{\partial}{\partial r}\ln\left(\frac{\Sigma(r,\theta) B^2(r,\theta) W^2(r,\theta)}{r^2K^2(r,\theta)(\partial N^2(r,\theta)/\partial r)^2}\right)\Biggr|_{r=r_0}.
\end{eqnarray}

If one fixes the arbitrary functions $T_0(\theta)$ and $\Psi_0(\theta)$, so that
\begin{eqnarray}\nonumber
T_0'(\theta)&=&-\dfrac{1}{2\kappa_g}\lim_{r\to r_0}\dfrac{1}{N^2(r,\theta)}\dfrac{\partial N^2(r,\theta)}{\partial \theta},
\\\nonumber
\Psi_0'(\theta)&=&-\dfrac{\Omega_H}{2\kappa_g}\lim_{r\to r_0}\dfrac{1}{N^2(r,\theta)}\dfrac{\partial N^2(r,\theta)}{\partial \theta},
\end{eqnarray}
the above coordinates approach (\ref{EFcoord}) at the horizon
\begin{equation}\label{EFcoord-regular}
\begin{array}{rcl}
 d\tau &=& dt+C(r,\theta)dr+\Order{r-r_0},\\ \PRDstyle{\\}
 d\phi &=& d\varphi+C(r,\theta)\dfrac{W(r,\theta)}{rK^2(r,\theta)}dr+\Order{r-r_0}.\PRDstyle{\\ \\}
\end{array}
\end{equation}

Using the coordinate transformation (\ref{regular-coord}) one can find explicitly the metric tensor in terms of the ingoing Eddington–Finkelstein-like coordinates.
The corresponding line element does not have coordinate singularity at the future horizon $r=r_0$. It can be given by Eq.~(\ref{EFmetric}) with additional terms depending on the coefficients of the expansion (\ref{regular-exp}), which do not change the results presented in Sec.~\ref{sec:BZ}.

In a similar way one can obtain the outgoing Eddington–Finkelstein-like coordinates penetrating the past horizon for the parametrized stationary axisymmetric black hole.

\end{document}